\documentstyle[preprint,aps]{revtex}
\begin{document}
\draft

\title{New phase diagram of Zn-doped CuGeO$_3$}

\author{Y. Sasago, N. Koide and K. Uchinokura}
\address{Department of Applied Physics, University of Tokyo, Bunkyo-ku, 
Tokyo 113, Japan}
\author{Michael C. Martin, M. Hase\thanks{Permanent address:
Institute of Physical and Chemical Research (RIKEN), Wako 351-01, Japan},
K. Hirota\thanks{Present address:
Department of Physics, Tohoku University, Sendai 980, Japan},
and G. Shirane}
\address{Department of Physics, Brookhaven National Laboratory, Upton, NY 
11973-5000}

\date{\today}
\maketitle

\begin{abstract}
A series of high-quality single crystals of Cu$_{1-x}$Zn$_x$GeO$_3$ have
been examined by neutron scattering techniques.  An antiferromagnetic
(AF) ordering is confirmed for samples with $x\geq 0.02$, in complete 
agreement with previous reports.  We show that the 
spin-Peierls (SP) phase transition persists to 6\% Zn, whereas 
previous magnetic susceptibility measurements reported a deterioration
of the SP transition above 2\% Zn.  We present some details of the
successive transitions upon lowering temperature into the spin-Peierls 
phase which is followed by a transition into an antiferromagnetically 
ordered phase.  Below the N\'eel temperature we observe the coexistence of 
the SP lattice dimerization and AF states. 

\end{abstract}
\pacs{PACS: } 

\narrowtext

\section{Introduction}

Shortly after the discovery of the inorganic spin-Peierls cuprate CuGeO$_3$
\cite{hase93}, a series of extensive studies were begun on systems where
Cu atoms were replaced with Zn 
\cite{hase93-2,hase95,oseroff,lussier,hase96} or Ge was replaced with Si
\cite{renard,poirier,regnault}.  
It is now well established that a new antiferromagnetically (AF)
ordered phase appears as shown in the phase diagram of Figure 
\ref{MagPhases} obtained on powder samples \cite{hase93-2}.  The spin-Peierls 
(SP) transition temperature is near 14K for the undoped material, decreases
in temperature with increased Zn concentration, and seemed 
to disappear around 2\% Zn \cite{hase93-2,hase96}; at 4\% Zn the magnetic 
susceptibility no longer shows a SP transition but only 
shows an AF transition with a N\'eel temperature of T$_{\rm N}\sim4$K 
\cite{hase95} (inset of Figure \ref{MagPhases}).  

Two recent neutron scattering reports have shown the existence of the AF
ordering with its associated superlattice peak at $(0\ 1\ {1\over 2})$ for
3.4\% Zn-doped \cite{lussier,hase96} and 0.7\% Si-doped \cite{regnault} 
CuGeO$_3$.
In the latter study, Regnault {\it et al.} showed the successive SP and
AF transitions with two separate branches of magnetic excitations below
T$_{\rm N}$ \cite{regnault}.  The coexistence of the SP and AF states was 
first demonstrated in their work.

The AF state in Zn-doped CuGeO$_3$ is certainly unusual.  Undoped, the low
dimensional chains in CuGeO$_3$ form a SP ground state with an accompanying
lattice dimerization below about 14 K.  The AF order is
induced when impurities are doped into this SP spin-singlet ground state.
The present paper presents preliminary neutron scattering results on 2\% and
4\% Zn-doped CuGeO$_3$ showing the SP and AF transitions and the interplay
between these two states.

\section{Experimental Details}

A series of relatively large ($\sim0.4$cm$^3$) Cu$_{1-x}$Zn$_x$GeO$_3$
single crystals were grown using the floating-zone method.  The nominal
values of $x$ for the crystals used in the present study are 0.02, 0.04, 
and 0.06 \cite{concentration}.  The space
group of CuGeO$_3$ is Pbmm (Pmma in standard orientation), with lattice
constants at room temperature of $a=4.81$\AA, $b=8.47$\AA, and $c=2.941$\AA.
The mosaic spread of these crystals were less than 0.3 degrees.
Neutron scattering measurements were carried out on the H7 and H8 beamlines
of the High Flux Beam Reactor at Brookhaven National Laboratory.  The
crystals were mounted in aluminum cans which were subsequently attached to
the cold finger of a cryostat.  The samples were aligned so as to place 
$(0\ k\ l)$ or $(h\ k\ h$) zones in the experimental scattering plane.
Incident neutrons with energies of 14.7meV were selected by a pyrolytic 
graphite (PG) monochromator and PG filters where used to eliminate higher-order
harmonics.  The beam was horizontally collimated typically with 
40'-40'-Sample-40'-80' in sequence from the reactor core to the detector.

\section{Phase Transitions}

We show the temperature dependence of the intensities of the two 
superlattice peaks at $({1\over 2}\ 6\ {1\over 2})$ due to the lattice 
dimerization which occurs below the SP transition, and at 
$(0\ 1\ {1\over 2})$ from the AF ordering for 
Cu$_{0.98}$Zn$_{0.02}$GeO$_3$ in Figure \ref{2Zn}.  The SP transition
temperature is reduced by about 3 degrees compared to the undoped material,
and the transition is broader indicating a possible range of T$_{\rm SP}$'s.
Below about 2K the AF superlattice peak is observed.
These two transition temperatures are in agreement with the previous powder
study by Hase {\it et al.} \cite{hase93-2}.
The SP dimerization superlattice peak intensity persists in the AF state 
with a very slight decrease in intensity below the N\'eel temperature 
(T$_{\rm N}$), showing the coexistence of these two states.  

Figure \ref{4Zn} again shows the $({1\over 2}\ 6\ {1\over 2})$ SP dimerization
peak and
the $(0\ 1\ {1\over 2})$ AF peak intensities as functions of temperature
now for a 4\% Zn-doped crystal.  A broadened and reduced T$_{\rm SP}$ is
again observed in this sample.  However the $({1\over 2}\ 6\ {1\over 2})$
peak has become noticeably weaker (note the right-hand scale).  The AF
peak intensity shows an onset giving a N\'eel temperature near 4K.  
In this sample we can
clearly observe a decrease in the SP superlattice peak intensity below
T$_{\rm N}$ (indicated in Fig. \ref{4Zn} by a dashed line).  This indicates
that while the two states are coexisting, the magnitude of the SP lattice
dimerization is affected by the onset of antiferromagnetism.  

Similar measurements of T$_{\rm SP}$ and T$_{\rm N}$ were performed on 
a 6\% Zn-doped CuGeO$_3$ crystal.  The results for all Zn-doped compounds
investigated are presented in the inset to Figure \ref{4Zn}.  In contrast
to the initial phase diagram determined by susceptibility measurements
(Figure \ref{MagPhases}), we have shown that the SP transition does not
go away upon doping with Zn, but instead remains at approximately 10K as the
dopant level is increased.  Furthermore the SP dimerization and AF ordered
states are observed
to coexist in all Zn-doped samples studied.  All Bragg peaks for the 2\% and
4\% Zn-doped crystals are resolution limited meaning that the SP and AF 
orderings are long-range in nature.

Figures \ref{2Zn} and \ref{4Zn} also show how the AF state becomes
more dominant as the Zn-doping increases.  The relative intensity of the
AF peak to the SP peak increases significantly with increasing
Zn-doping.  Comparing the scattering intensities for the two crystals
presented we can see that for the 4\% Zn sample the
AF scattering has increased by approximately two-fold over the 2\% Zn sample
(indicating an increase in the magnetic moment of the AF state), 
while the SP dimerization
scattering intensity has decreased.  In the 4\% Zn sample we estimate
the zero temperature magnetic moment to be $\mu \approx
0.2\mu_{\rm B}$; this is quite close to the value
($\mu \approx 0.22\mu_{\rm B}$) reported by
Hase {\it et al.} \cite{hase96}.
The decrease in the intensity of the SP superlattice peak can be used to
estimate the decrease in the
atomic displacement $\delta$ from the SP dimerization in pure CuGeO$_3$.
Using the observed form factor of the (0\ 2\ 1) Bragg peak as a
reference and comparing the previously measured pure CuGeO$_3$
results (here we will denote the
atomic displacement of a sample with $x$ Zn dopant as $\delta_x$) we
find that $\delta_{0.04}\approx {2\over{3}}\delta_{0}$.

\section{Discussion}

The most interesting result from the study of doped CuGeO$_3$ is the 
coexistence of the SP lattice dimerization and the N\'eel state at low 
temperatures.
This was first reported for a 0.7\% Si-doped sample by Regnault {\it et al.}
\cite{regnault} and is reported in the present paper for a wide range of Zn 
doping.  Usually two ordered phases of these types are mutually exclusive.
Recently Fukuyama {\it et al.} \cite{fukuyama} proposed a theoretical 
model of disorder-induced antiferromagnetic long range order in a
spin-Peierls system.  Some key features of the theory appear to be 
realized in the current neutron data of the Zn-doped system.  More 
quantitative data are needed for a concrete comparison with this theory;
accurate and reliable determinations of the two order parameters with
increasing $x$ and eventual line broadening of the SP and AF superlattice
peaks at higher dopant concentrations are required.

The spectral shapes of magnetic excitations are also of vital interest.
This phase of our investigations is just being undertaken, and some 
examples of interesting results are shown in Figure \ref{magpeak}.  The
magnetic excitations of Cu$_{0.98}$Zn$_{0.02}$GeO$_3$ and 
Cu$_{0.96}$Zn$_{0.04}$GeO$_3$ are essentially 
unchanged from the undoped crystals, remaining sharp and well defined.
This result is in sharp contrast to the overdamped magnetic excitations for
an $x=0.04$ sample reported recently by Hase {\it et al.} \cite{hase96note}.
This difference is presumably due to the improved quality of the present
crystals though we lack microscopic probes to demonstrate the nature of
the improvement.

Regnault {\it et al.} \cite{regnault} reported differences in the magnetic
excitations of the two phases including a new AF branch at low energies
(or low $q$) and the shift of spectral weight.  We do observe some significant
shifts as demonstrated in the lower panel of Fig. \ref{magpeak}.  However 
in contrast to the report of Regnault {\it et al.} we have not observed sharp 
AF excitations.

The results presented here, together with the Si-doped data of Regnault 
{\it et al.} \cite{regnault}
clearly demonstrate the coexistence of antiferromagnetic ordering and 
spin-Peierls lattice dimerization.  Research is 
continuing to further clarify any changes that occur as the SP phase develops
an AF ordering.

\acknowledgments
We would like to thank R.J. Birgeneau, Guillermo Castilla, Vic Emery, and
H. Fukuyama for stimulating discussions.  
This work was supported in part by the U.S.- Japan Cooperative Research 
Program on Neutron Scattering, and 
NEDO (New Energy and Industrial Technology Development Organization) 
International Joint Research Grant.
Research at Brookhaven National Laboratory was supported by the Division of 
Materials Research at the U.S. Department of Energy, under contract No. 
DE-AC02-76CH00016.

\begin{figure}
\caption{The previously reported phase diagram for Cu$_{1-x}$Zn$_x$GeO$_3$
as deduced from magnetic susceptibility measurements of powders 
\protect{\cite{hase93-2}}.  The inset shows the susceptibility measurement of
an $x=0.04$ single crystal \protect{\cite{hase95}}.}
\label{MagPhases}
\end{figure}

\begin{figure}
\caption{Intensities of the (0.5 6 0.5) SP superlattice peak and the (0 1 0.5)
AF superlattice peak as a function of temperature for a 2\% Zn-doped sample
showing the onset of the SP state followed by the onset of the AF state.
The SP lattice dimerization superlattice peak was followed down to 1.3K and 
shows a small decrease below 2K.}
\label{2Zn}
\end{figure}

\begin{figure}
\caption{Intensities of the SP and AF superlattice peaks as functions of 
temperature for a 4\% Zn-doped crystal.  The intensity of the SP lattice 
dimerization peak is seen to decrease below T$_{\rm N}$, 
however the states are clearly coexisting.  The inset
shows T$_{\rm SP}$ and T$_{\rm N}$ measured on samples of 0, 2, 4, and 6
percent Zn-doped crystals.  The two solid lines reproduce the phase diagram
of Hase {\it et al.} (Fig \protect{\ref{MagPhases}}) for comparison.}
\label{4Zn}
\end{figure}

\begin{figure}
\caption{Magnetic excitation spectra for a 2\% (top panel) and 4\% (bottom)
Zn-doped CuGeO$_3$ crystals.  The symbols are the data and the lines are fits 
to two Gaussians.  The open triangles and solid line fit correspond to data
taken at T $<$ T$_{\rm N}$ and the solid diamonds with a dashed line fit are
for T$_{\rm N} <$ T $<$ T$_{\rm SP}$.}
\label{magpeak}
\end{figure}


\begin{references}
\bibitem{concentration}The concentration of Zn in the 4\% Zn-doped sample 
used in Ref. \protect{\cite{hase96}} was
determined to be 0.034 by electron-probe microanalysis.  A similar test may 
systematically shift the current values of $x$ to somewhat lower values.
\end{references}
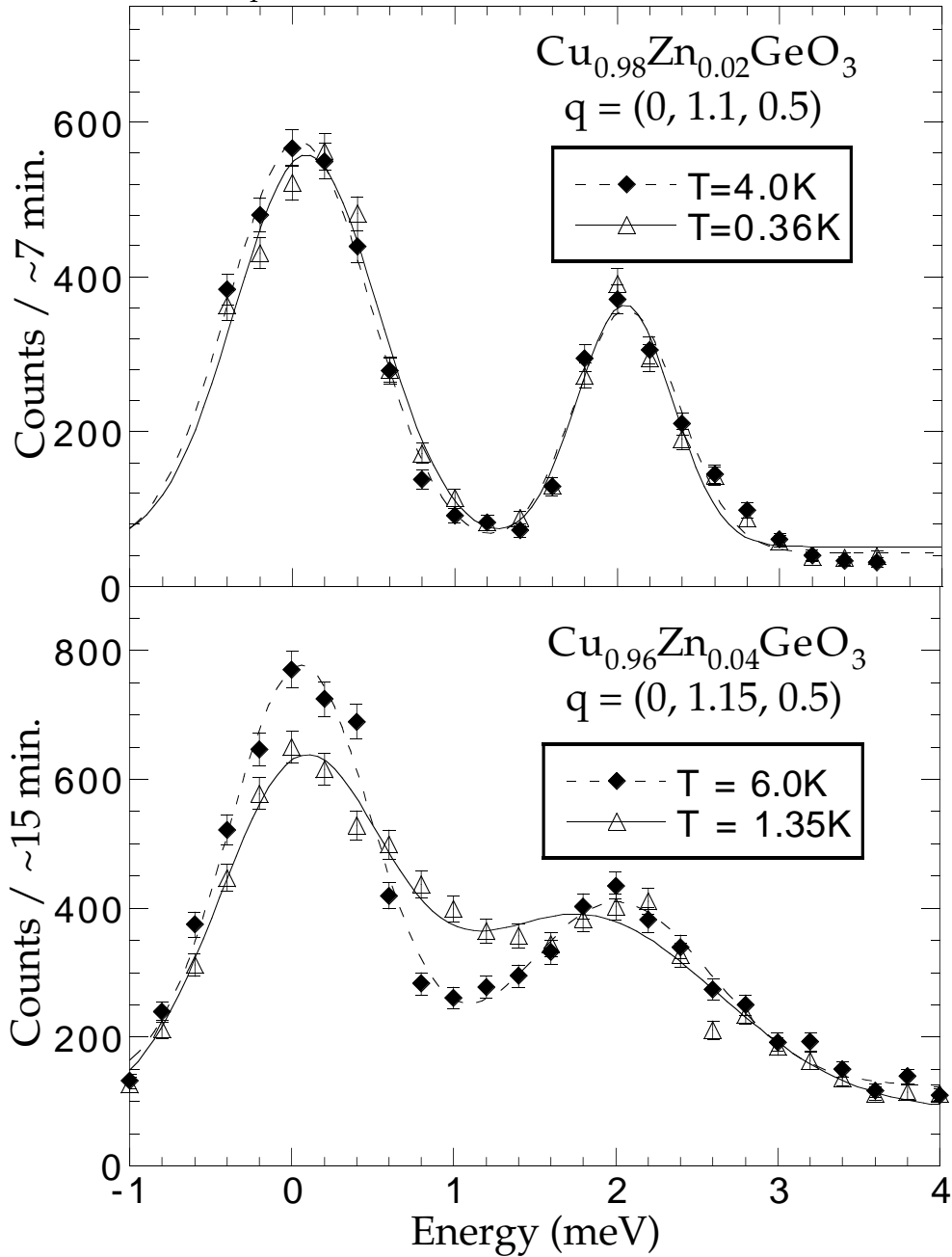
\end{document}